\documentclass[aps,prc,twocolumn,showpacs,preprintnumbers,amsmath,amssymb,
epsf,superscriptaddress,groupedaddress,nofootinbib,tightenlines]{revtex4}

\usepackage{graphicx}
\usepackage{amsmath}
\usepackage{bm}

\def\beqn{\begin{eqnarray}}
\def\eeqn{\end{eqnarray}}
\def\barr{\begin{array}}
\def\earr{\end{array}}
\def\btab{\begin{tabular}}
\def\etab{\end{tabular}}
\def\bite{\begin{itemize}}
\def\eite{\end{itemize}}
\def\bcen{\begin{center}}
\def\ecen{\end{center}}

\def\eq{\begin{equation}}
\def\ee{\end{equation}}
\def\nn{\nonumber}

\def\kdagger{K\hspace{-0.3cm}/\;}
\def\ndagger{n\hspace{-0.2cm}/}

\def\xidagger{\xi\hspace{-0.18cm}/}

\def\keldagger{k\hspace{-0.2cm}/}

\def\q2dagger{q_2\hspace{-0.35cm}/\;}

\begin{document}
\title{Doubly virtual Compton scattering and the beam normal spin asymmetry}
\author{Mikhail Gorchtein}
\affiliation{California Institute of Technology, Pasadena, CA 91125, USA}
\email{gorshtey@caltech.edu}
\begin{abstract}
We construct an invariant basis for Compton scattering with two virtual 
photons (VVCS). 
The  basis tensors are chosen to be gauge invariant and orthogonal 
to each other. The properties of the corresponding 18 invariant amplitudes 
are studied in detail. We consider the special case of elastic VVCS with 
the virtualities of the initial and final photons equal. The invariant 
basis for VVCS in this orthogonal form does not exist in the literature. 
We furthermore use this VVCS tensor for a calculation of the beam normal spin 
asymmetry in the forward kinematics. For this, we relate the invariant 
amplitudes to the helicity amplitudes of the VVCS reaction. The imaginary parts 
of the latter are related to the inclusive cross section by means of the 
optical theorem. We use the phenomenological value of the transverse 
cross section $\sigma_T\sim0.1$ mbarn and the Callan-Gross relation which 
relates the longitudinal cross section $\sigma_L$ to the transverse one.
The result of the calculation agrees with an existing calculation, and predicts
 negative values of the asymmetry $B_n$ of order of 5 ppm in the 
energy range from 3 to 45 GeV and for very forward angles.
\end{abstract}
\maketitle
\section{Introduction}
Over recent years, interest in Compton scattering with two virtual 
(typically, space-like) photons has arisen and has been constantly growing. 
This interest is based on significant progress in precision
electron-proton scatering experimental technique. Polarization transfer 
data for elastic electron-proton scattering \cite{gegm_poltransfer} showed 
that the extracted values of the electromagnetic form factors obtained with 
this technique differ significantly from the Rosenbluth separation data 
\cite{gegm_rosenbluth}. It has been pointed out that these two data sets may 
be reconciled by the proper inclusion of higher order contributions 
in the electromagnetic coupling constant $\alpha_{em}$, namely the two photon 
exchange contribution \cite{marcguichon}. 
Several model calculations using the elastic form factors only 
\cite{blunden1}, the $\Delta(1232)$ resonance \cite{blunden2}, and the partonic 
picture \cite{marc_etal}, show that the contribution of the two photon exchange 
exhibits the correct trend which may be enough to explain the discrepancy of 
the experimental results.
On the other hand, a different class of observables has become 
experimentally accessible 
in the framework of parity-violating electron-proton 
scattering. These new observables involve the spin orientation normal to the 
reaction plane and have been shown to be directly related to the imaginary part 
of the elastic electron-proton scattering amplitude. Since one photon exchange 
does not lead to an imaginary part, the leading order contribution comes from 
the exchange of at least two photons \cite{derujula}. Since the exchange of 
three or more photons is suppressed by powers of the small coupling constant 
$\alpha_em$, a measurement of such an observable probably provides the 
cleanest way to access Compton scattering with two virtual photons. At present, 
several experimental data points exist in different kinematical regimes 
\cite{bn_exp}. 
On the theoretical side, significant progress has been made towards 
quantitative understanding of the experimental data \cite{bn_theo}. \\
\indent
These experimental and theoretical studies require a description of  
Compton scattering with two virtual photons. The general form of the 
explicitly gauge invariant basis for VVCS was developed 
in the 1970's by Tarrach \cite{tarrach}. The invariant amplitudes of Tarrach 
have the advantage that they don't have kinematical singularities nor 
constraints, and thus can be used for a dispersion calculation.
However, they were never used for a realistic calculation. 
One of the reasons is 
that, at that time, the VVCS reaction was of pure theoretical interest 
(apart from its forward limit which is used in DIS and described by the 
nucleon structure functions rather than the invariant amplitudes of Tarrach). 
The other reason is the technical difficulty in using this basis 
in a real calculation, which is due to the non-orthogonality of the basis 
tensors. Another approach was used by Prange for real Compton scattering 
\cite{prange} and then generalized to Compton scattering with a virtual 
initial photon by Berg and Lindner \cite{berglindner}. This method consists 
in defining the invariant Compton basis in terms of gauge invariant basis 
tensors which are orthogonal to each other. In this work, we generalize this 
approach to the case when the outgoing photon is also virtual. In 
section \ref{sec:vvcs_tensor}, we introduce the VVCS tensor in such a form, 
and study the properties of the invariant amplitudes. We also consider the 
special case of elastic VVCS with initial and outgoing photons 
of equal virtuality. In  the literature, the invariant Compton tensor for 
this reaction exists only for the forward case. We obtain here the result 
for the general case. In section \ref{sec:appl_bn}, we use the introduced 
from of the Compton tensor for a practical calculation of the beam normal spin 
asymmetry $B_n$ in the forward kinematics. 
\section{Doubly VCS tensor basis}
\label{sec:vvcs_tensor}
In this section, the doubly virtual Compton scattering (VVCS) tensor will be 
introduced and the properties of the invariant VVCS amplitudes will be 
investigated. We start by reviewing existing developments on invariant Compton 
tensors. A general Lorentz and gauge-invariant VVCS tensor was constructed 
by Tarrach 
\cite{tarrach} with the corresponding amplitudes free from kinematical 
singularities and constraints. However, it appears 
problematic to use this tensor 
basis for practical calculations. Therefore, we turn to the method used by 
Prange for RCS \cite{prange} and extended to VCS (with only one virtual 
photon) by Berg and Lindner \cite{berglindner}. We consider a scattering 
process $\gamma^*(q_1)+N(p)\to\gamma^*(q_2)+N(p')$ and define the four-vectors:
\beqn
P&=&\frac{p+p'}{2}\nn\\
\tilde{K}&=&\frac{q_1+q_2}{2}\nn\\
L&=&\frac{q_1-q_2}{2}\,=\,\frac{p'-p}{2}.
\eeqn
The method of Ref.\cite{prange} for RCS amounts to defining two 
orthogonal vectors,
\beqn
P'^\mu&=&P^\mu-\frac{P\cdot\tilde{K}}{\tilde{K}^2}\tilde{K}^\mu,\nn\\
n^\mu&=&\varepsilon^{\mu\alpha\beta\gamma}P'_\alpha\tilde{K}_\beta L_\gamma,
\eeqn
where we use the convention $\varepsilon^{0123}=-1$. The four vectors 
$P',n,L'\tilde{K}$ form an orthogonal basis. Out of these vectors, 
orthogonal and explicitly gauge invariant tensors can be constructed. 
The six independent 
Lorentz-invariant structures which are needed to describe RCS in the most 
general case are 
\cite{prange}:
\beqn
{M}^{\mu\nu}_{RCS}&=&\bar{N}'
\left\{
\frac{P'^\mu P'^\nu}{P'^2}(B_1\,+\,\kdagger B_2)\;+\;
\frac{n^\mu n^\nu}{n^2}(B_3\,+\,\kdagger B_4)\right.\nn\\
&&\;\;\;\;\;\;
\;+\frac{P'^\mu n^\nu\,-\,n^\mu P'^\nu}{P'^2n^2}\,i\gamma_5 B_7\nn\\
&&\;\;\;\;\;\;
\left.\;+\frac{P'^\mu n^\nu\,+\,n^\mu P'^\nu}{P'^2n^2}\,\ndagger B_6
\right\}\,N,
\label{eq:rcstensor}
\eeqn

We note that the terms 
$\sim\frac{P'^\mu n^\nu\,+\,n^\mu P'^\nu}{P'^2n^2}\,i\gamma_5$ and 
$\frac{P'^\mu n^\nu\,-\,n^\mu P'^\nu}{P'^2n^2}\,\ndagger$ are ruled out by 
time-reversal invariance.
If we allow for one of the photons (normally, the initial one) to be virtual, 
this restriction is relaxed because the initial and final states are not 
identical anymore. The
method should be modified as follows \cite{berglindner}: since the condition 
$L\cdot\tilde{K}=0$ is no longer valid, we introduce
\beqn
L'^\mu&=&L^\mu-\frac{L\cdot\tilde{K}}{\tilde{K}^2}\tilde{K}^\mu,\nn\\
P'^\mu&=&P^\mu-\frac{P\cdot\tilde{K}}{\tilde{K}^2}\tilde{K}^\mu 
-\frac{P\cdot L'}{L'^2}L'^\mu,\nn\\
n^\mu&=&\varepsilon^{\mu\alpha\beta\gamma}P'_\alpha\tilde{K}_\beta L'_\gamma,
\eeqn
which are orthogonal to each other and to the vector $\tilde{K}$. 
Furthermore, in order to take into account the longitudinal polarization of 
the virtual photon, a gauge invariant vector is introduced, 
\beqn
\tilde{K}'^\mu&=&\tilde{K}^\mu
-\frac{q_1\cdot\tilde{K}}{q_1\cdot L'}L'^\mu,
\eeqn
which obeys $q_1\cdot\tilde{K}'=0$. With these modifications, the most 
general VCS tensor basis takes the form \cite{berglindner}:
\beqn
{M}^{\mu\nu}_{VCS}&=&\bar{N}'
\left\{
\frac{P'^\mu P'^\nu}{P'^2}(B_1\,+\,\kdagger B_2)\;+\;
\frac{n^\mu n^\nu}{n^2}(B_3\,+\,\kdagger B_4)\right.\nn\\
&&\;\;\;\;\;\;
\;+\frac{P'^\mu n^\nu\,+\,n^\mu P'^\nu}{P'^2n^2}\,
(i\gamma_5 B_5\,+\,\ndagger B_6)\nn\\
&&\;\;\;\;\;\;
\;+\frac{P'^\mu n^\nu\,-\,n^\mu P'^\nu}{P'^2n^2}
\,(i\gamma_5 B_7\,+\,\ndagger B_8)\nn\\
&&\;\;\;\;\;\;
\;+\frac{\tilde{K}'^\mu P'^\nu}{P'^2\tilde{K}^2}
\,(B_9\,+\,\kdagger B_{10})\nn\\
&&\;\;\;\;\;\;
\left.\;+\frac{\tilde{K}'^\mu n^\nu}{n^2\tilde{K}^2}
\,(i\gamma_5 B_{11}\,+\,\ndagger B_{12})
\right\}\,N,
\label{eq:vcstensor}
\eeqn

The Compton tensor in this form, for the case where both photons are virtual 
does not exist in the literature. We follow here the scheme proposed in 
\cite{berglindner} and generalize the invariant Compton tensor for the VVCS 
case. For this, we introduce another gauge invariant vector with respect to 
the outgoing virtual photon $q_2$,
\beqn
\tilde{K}''^\nu&=&\tilde{K}^\nu
-\frac{q_2\cdot\tilde{K}}{q_2\cdot L'}L'^\nu.
\eeqn

We finally obtain the 18 different structures that contain all the information 
about Compton scattering with two virtual photons:
\beqn
{M}^{\mu\nu}_{VVCS}&=&\bar{N}'
\left\{
\frac{P'^\mu P'^\nu}{P'^2}(B_1\,+\,\kdagger B_2)\;+\;
\frac{n^\mu n^\nu}{n^2}(B_3\,+\,\kdagger B_4)\right.\nn\\
&&\;\;\;\;\;\;
\;+\frac{P'^\mu n^\nu\,+\,n^\mu P'^\nu}{P'^2n^2}\,
(i\gamma_5 B_5\,+\,\ndagger B_6)\nn\\
&&\;\;\;\;\;\;
\;+\frac{P'^\mu n^\nu\,-\,n^\mu P'^\nu}{P'^2n^2}
\,(i\gamma_5 B_7\,+\,\ndagger B_8)\nn\\
&&\;\;\;\;\;\;
\;+\frac{\tilde{K}'^\mu P'^\nu}{P'^2\tilde{K}^2}
\,(B_9\,+\,\kdagger B_{10})\nn\\
&&\;\;\;\;\;\;
\;+\frac{\tilde{K}'^\mu n^\nu}{n^2\tilde{K}^2}
\,(i\gamma_5 B_{11}\,+\,\ndagger B_{12})\nn\\
&&\;\;\;\;\;\;
\;+\frac{P'^\mu\tilde{K}''^\nu}{P'^2\tilde{K}^2}
\,(B_{13}\,+\,\kdagger B_{14})\nn\\
&&\;\;\;\;\;\;
\;+\frac{n^\mu\tilde{K}''^\nu}{n^2\tilde{K}^2}
\,(i\gamma_5 B_{15}\,+\,\ndagger B_{16})\nn\\
&&\left.\;\;\;\;\;\;
\;+\frac{\tilde{K}'^\mu\tilde{K}''^\nu}{\tilde{K}^2}
\,(B_{17}\,+\,\kdagger B_{18})
\right\}\,N.
\label{eq:vvcstensor}
\eeqn

The invariant amplitudes $B_i$ are functions of the invariants:
\beqn
B_i\;=\;B_i(q_1^2,q_2^2,q_1\cdot q_2,P\cdot\tilde{K}).
\eeqn

We next consider the behaviour of the invariant amplitudes under photon and 
nucleon crossing.
\subsection{Nucleon crossing}
This transformation is a combination of partial parity transformation and 
charge conjugation:$\Big[\gamma^*(q_1)+N(p)\to\gamma^*(q_2)+N(p')\Big]\;\rightarrow\;\Big[\gamma^*(q_1)+\bar{N}(-p')\to\gamma^*(q_2)+\bar{N}(-p)\Big].$
We note that this is an unphysical transformation since only the hadronic 
part of the kinematics is transformed, while the photons remain unchanged.
This transformation, however, relates two reactions which should have the same 
structure. Under nucleon crossing, we have:
\beqn
P&\to&-P,\nn\\
n&\to&-n,\nn\\
\tilde{K}&\to&\tilde{K},\nn\\
L'&\to&L',\nn\\
\tilde{K}'&\to&\tilde{K}',\nn\\
\tilde{K}''&\to&\tilde{K}'',\nn\\
\kdagger&\to&{\cal{C}}^\dagger\kdagger{\cal{C}}=-\kdagger,\nn\\
\ndagger&\to&-{\cal{C}}^\dagger\ndagger{\cal{C}}=\ndagger,\nn\\
\gamma_5&\to&{\cal{C}}^\dagger\gamma_5{\cal{C}}=\gamma_5.
\eeqn

Hence, nucleon crossing transforms the VVCS tensor as 
\beqn
^N\!\tilde{M}^{\mu\nu}_{VVCS}&=&\bar{N}
\left\{
\frac{P'^\mu P'^\nu}{P'^2}
(\,^N\!\tilde{B}_1\,-\,\kdagger\,^N\!\tilde{B}_2)\right.\\
&&\;\;\;\;\;\;
\;+\frac{n^\mu n^\nu}{n^2}
(\,^N\!\tilde{B}_3\,-\,\kdagger\,^N\!\tilde{B}_4)\nn\\
&&\;\;\;\;\;\;
\;+\frac{P'^\mu n^\nu\,+\,n^\mu P'^\nu}{P'^2n^2}\,
(i\gamma_5\,^N\!\tilde{B}_5\,+\,\ndagger\,^N\!\tilde{B}_6)\nn\\
&&\;\;\;\;\;\;
\;+\frac{P'^\mu n^\nu\,-\,n^\mu P'^\nu}{P'^2n^2}
\,(i\gamma_5\,^N\!\tilde{B}_7\,+\,\ndagger\,^N\!\tilde{B}_8)\nn\\
&&\;\;\;\;\;\;
\;-\frac{\tilde{K}'^\mu P'^\nu}{P'^2\tilde{K}^2}
\,(\,^N\!\tilde{B}_9\,-\,\kdagger\,^N\!\tilde{B}_{10})\nn\\
&&\;\;\;\;\;\;
\;-\frac{\tilde{K}'^\mu n^\nu}{n^2\tilde{K}^2}
\,(i\gamma_5\,^N\!\tilde{B}_{11}\,+\,\ndagger\,^N\!\tilde{B}_{12})\nn\\
&&\;\;\;\;\;\;
\;-\frac{P'^\mu\tilde{K}''^\nu}{P'^2\tilde{K}^2}
\,(\,^N\!\tilde{B}_{13}\,-\,\kdagger\,^N\!\tilde{B}_{14})\nn\\
&&\;\;\;\;\;\;
\;-\frac{n^\mu\tilde{K}''^\nu}{n^2\tilde{K}^2}
\,(i\gamma_5\,^N\!\tilde{B}_{15}\,+\,\ndagger\,^N\!\tilde{B}_{16})\nn\\
&&\left.\;\;\;\;\;\;
\;+\frac{\tilde{K}'^\mu\tilde{K}''^\nu}{\tilde{K}^2}
\,(\,^N\!\tilde{B}_{17}\,-\,\kdagger\,^N\!\tilde{B}_{18})
\right\}\,N'.\nn
\label{eq:ncrossing}
\eeqn
\indent
In the above equation, $^N\!\tilde{B}_i$ stands for the nucleon crossing 
transform of the amplitude $B_i$. 
For the invariant amplitudes, nucleon crossing amounts to the substitution 
$P\cdot\tilde{K}\to -P\cdot\tilde{K}$, while the other arguments remain 
unchanged.
We next list the behavior of the invariant amplitudes $B_i$ under nucleon 
crossing:
\beqn
{B}_{i}
(q_1^2,q_2^2,q_1\cdot q_2,-P\cdot\tilde{K})
&=&+{B}_{i}(q_1^2,q_2^2,q_1\cdot q_2,P\cdot\tilde{K})\nn\\
{\rm for}\;i&=&1,3,5,6,7,8,10,14,17\nn\\
{B}_{i}
(q_1^2,q_2^2,q_1\cdot q_2,-P\cdot\tilde{K})
&=&-{B}_{i}(q_1^2,q_2^2,q_1\cdot q_2,P\cdot\tilde{K})\nn\\
{\rm for}\;i&=&2,4,9,11,12,13,15,16,18\nn\\
&&
\label{eq:bi_ncrossing}
\eeqn

\subsection{Photon crossing}
Photon crossing relates two reactions: $\Big[\gamma^*(q_1)+N(p)\to\gamma^*(q_2)+N(p')\Big]\;\rightarrow\;\Big[\gamma^*(-q_2)+N(p)\to\gamma^*(-q_1)+N(p')\Big].$
In the general case of $q_1^2\neq q_2^2$, the Compton amplitude must not be 
invariant under photon crossing. It is, however, a relevant symmetry in the 
elastic case $q_1^2= q_2^2$.
Under photon crossing we have:
\beqn
\tilde{K}&\to&-\tilde{K},\nn\\
\mu&\leftrightarrow&\nu\nn\\
P&\to&P,\nn\\
P'&\to&P',\nn\\
L'&\to&L',\nn\\
n&\to&-n,\nn\\
\tilde{K}'&\to&-\tilde{K}'',\nn\\
\tilde{K}''&\to&-\tilde{K}'.
\eeqn

Hence, photon crossing transforms the VVCS tensor as 
\beqn
^\gamma\!\tilde{M}^{\mu\nu}_{VVCS}&=&\bar{N}'
\left\{
\frac{P'^\mu P'^\nu}{P'^2}
(\,^\gamma\!\tilde{B}_1\,-\,\kdagger\,^\gamma\!\tilde{B}_2)\right.\\
&&\;\;\;\;\;\;
\;+\frac{n^\mu n^\nu}{n^2}
(\,^\gamma\!\tilde{B}_3\,-\,\kdagger\,^\gamma\!\tilde{B}_4)\nn\\
&&\;\;\;\;\;\;
\;-\frac{P'^\mu n^\nu\,+\,n^\mu P'^\nu}{P'^2n^2}\,
(i\gamma_5\,^\gamma\!\tilde{B}_5\,-\,\ndagger\,^\gamma\!\tilde{B}_6)\nn\\
&&\;\;\;\;\;\;
\;+\frac{P'^\mu n^\nu\,-\,n^\mu P'^\nu}{P'^2n^2}
\,(i\gamma_5\,^\gamma\!\tilde{B}_7\,-\,\ndagger\,^\gamma\!\tilde{B}_8)\nn\\
&&\;\;\;\;\;\;
\;-\frac{P'^\mu\tilde{K}''^\nu}{P'^2\tilde{K}^2}
\,(\,^\gamma\!\tilde{B}_{9}\,-\,\kdagger\,^\gamma\!\tilde{B}_{10})\nn\\
&&\;\;\;\;\;\;
\;+\frac{n^\mu\tilde{K}''^\nu}{n^2\tilde{K}^2}
\,(i\gamma_5\,^\gamma\!\tilde{B}_{11}
\,+\,\ndagger\,^\gamma\!\tilde{B}_{12})\nn\\
&&\;\;\;\;\;\;
\;-\frac{\tilde{K}'^\mu P'^\nu}{P'^2\tilde{K}^2}
\,(\,^\gamma\!\tilde{B}_{13}\,-\,\kdagger\,^\gamma\!\tilde{B}_{14})\nn\\
&&\;\;\;\;\;\;
\;+\frac{\tilde{K}'^\mu n^\nu}{n^2\tilde{K}^2}
\,(i\gamma_5\,^\gamma\!\tilde{B}_{15}
\,+\,\ndagger\,^\gamma\!\tilde{B}_{16})\nn\\
&&\left.\;\;\;\;\;\;
\;+\frac{\tilde{K}'^\mu\tilde{K}''^\nu}{\tilde{K}^2}
\,(\,^\gamma\!\tilde{B}_{17}\,-\,\kdagger\,^\gamma\!\tilde{B}_{18})
\right\}\,N'.\nn
\label{eq:phcrossing}
\eeqn
\indent
In the above equation, $^\gamma\!\tilde{B}_i$ stands for the nucleon crossing 
transform of the amplitude $B_i$. Apart from the substitution 
$P\cdot\tilde{K}\to -P\cdot\tilde{K}$, photon crossing interchanges the 
virtualities of the photons as arguments of the invariant amplitudes, 
$q_1^2\leftrightarrow q_2^2$. We find following behaviour under photon 
crossing:
\beqn
{B}_{i}
(q_2^2,q_1^2,q_1\cdot q_2,-P\cdot\tilde{K})
&=&+{B}_{i}(q_1^2,q_2^2,q_1\cdot q_2,P\cdot\tilde{K})\nn\\
{\rm for}\;i&=&1,3,6,7,17\nn\\
{B}_{i}
(q_2^2,q_1^2,q_1\cdot q_2,-P\cdot\tilde{K})
&=&-{B}_{i}(q_1^2,q_2^2,q_1\cdot q_2,P\cdot\tilde{K})\nn\\
{\rm for}\;i&=&2,4,5,8,18\nn\\
{B}_{9}
(q_2^2,q_1^2,q_1\cdot q_2,-P\cdot\tilde{K})
&=&-{B}_{13}(q_1^2,q_2^2,q_1\cdot q_2,P\cdot\tilde{K})\nn\\
{B}_{10}
(q_2^2,q_1^2,q_1\cdot q_2,-P\cdot\tilde{K})
&=&{B}_{14}(q_1^2,q_2^2,q_1\cdot q_2,P\cdot\tilde{K})\nn\\
{B}_{13}
(q_2^2,q_1^2,q_1\cdot q_2,-P\cdot\tilde{K})
&=&-{B}_{9}(q_1^2,q_2^2,q_1\cdot q_2,P\cdot\tilde{K})\nn\\
{B}_{14}
(q_2^2,q_1^2,q_1\cdot q_2,-P\cdot\tilde{K})
&=&{B}_{10}(q_1^2,q_2^2,q_1\cdot q_2,P\cdot\tilde{K})\nn\\
{B}_{11}
(q_2^2,q_1^2,q_1\cdot q_2,-P\cdot\tilde{K})
&=&{B}_{15}(q_1^2,q_2^2,q_1\cdot q_2,P\cdot\tilde{K})\nn\\
{B}_{12}
(q_2^2,q_1^2,q_1\cdot q_2,-P\cdot\tilde{K})
&=&-{B}_{16}(q_1^2,q_2^2,q_1\cdot q_2,P\cdot\tilde{K})\nn\\
{B}_{15}
(q_2^2,q_1^2,q_1\cdot q_2,-P\cdot\tilde{K})
&=&{B}_{11}(q_1^2,q_2^2,q_1\cdot q_2,P\cdot\tilde{K})\nn\\
{B}_{16}
(q_2^2,q_1^2,q_1\cdot q_2,-P\cdot\tilde{K})
&=&-{B}_{12}(q_1^2,q_2^2,q_1\cdot q_2,P\cdot\tilde{K})\nn\\
&&
\label{eq:bi_phcrossing}
\eeqn

We now combine photon and nucleon crossing, which corresponds to 
applying a $CP$-transformation to the full reaction. Combining 
Eqs.(\ref{eq:bi_ncrossing},\ref{eq:bi_phcrossing}), we obtain  
\beqn
B_{5,8},\,B_9-B_{13},\,B_{10}-B_{14},\,&&\\
B_{11}+B_{15},\,B_{12}+B_{16}
&\sim&(q_1^2-q_2^2)^{2n+1},\nn\\
B_{1,2,3,4,6,7,17,18},\,B_9+B_{13},\,B_{10}+B_{14},\,&&\nn\\
B_{11}-B_{15},\,B_{12}-B_{16}
&\sim&(q_1^2\pm q_2^2)^{2n},\nn
\eeqn
with $n=0,1,2,\dots$ \footnote{We do not expect any negative powers of 
$q_1^2-q_2^2$ since it would lead to an unphysical singularity.}

As a consequence, the VVCS tensor simplifies in the elastic case (EVVCS), 
$q_1^2=q_2^2$:
\beqn
{M}^{\mu\nu}_{EVVCS}&=&\bar{N}'
\left\{
\frac{P'^\mu P'^\nu}{P'^2}(B_1\,+\,\kdagger B_2)\right.\nn\\
&&\;\;\;\;\;\;\;+
\frac{n^\mu n^\nu}{n^2}(B_3\,+\,\kdagger B_4)\nn\\
&&\;\;\;\;\;\;
\;+\frac{P'^\mu n^\nu\,+\,n^\mu P'^\nu}{P'^2n^2}\,
\ndagger B_6\nn\\
&&\;\;\;\;\;\;
\;+\frac{P'^\mu n^\nu\,-\,n^\mu P'^\nu}{P'^2n^2}
\,i\gamma_5 B_7\nn\\
&&\;\;\;\;\;\;
\;+\frac{\tilde{K}'^\mu P'^\nu + P'^\mu\tilde{K}''^\nu}{P'^2\tilde{K}^2}
\,(B_9\,+\,\kdagger B_{10})\nn\\
&&\;\;\;\;\;\;
\;+\frac{\tilde{K}'^\mu n^\nu-n^\mu\tilde{K}''^\nu}{n^2\tilde{K}^2}
\,i\gamma_5 B_{11}\nn\\
&&\;\;\;\;\;\;
\;+\frac{\tilde{K}'^\mu n^\nu+n^\mu\tilde{K}''^\nu}{n^2\tilde{K}^2}
\,\ndagger B_{12}\nn\\
&&\left.\;\;\;\;\;\;
\;+\frac{\tilde{K}'^\mu\tilde{K}''^\nu}{\tilde{K}^2}
\,(B_{17}\,+\,\kdagger B_{18})
\right\}\,N.
\label{eq:evvcstensor}
\eeqn

We finally investigate the RCS limit, i.e. let the virtualities of the photons 
go to zero. In this case, 
\beqn
\tilde{K}'^\mu&\to&q_1^\mu\nn\\
\tilde{K}''^\nu&\to&q_2^\nu.
\eeqn

In an observable, the Compton tensor is contracted with the photon 
polarization vectors, $\varepsilon_\lambda(q_1)\cdot\tilde{K}'\to\varepsilon_\lambda(q_1)\cdot q_1=0$.
Though the amplitudes $B_{9,10,11,12,17,18}$ should not necessarily vanish in 
this limit, the corresponding terms do not contribute to any observables. 
Therefore, at the real photon point we recover the old RCS tensor of Prange 
of Eq.(\ref{eq:rcstensor}).

It should be stressed that doubly VCS tensors for inelastic 
($q_1^2\neq q_2^2$) and elastic ($q_1^2=q_2^2$) cases in the form of Eqs. 
(\ref{eq:vvcstensor},\ref{eq:evvcstensor}) do not exist in the literature.

\section{Application: Beam Normal Spin Asymmetry}
\label{sec:appl_bn}
We consider elastic electron-proton scattering 
$e^-(k)+N(p)\to e^-(k')+N(p')$, with the electron beam polarized
 normally to the scattering plane with the polarization vector defined as 
\beqn
\xi^\mu
\;=\;
\left(0,
\frac{\vec{k}\times\vec{k'}}{|\vec{k}\times\vec{k'}|}
\right).
\eeqn
\indent
The single spin asymmetry defined as 
$B_n=\frac{\sigma_\uparrow-\sigma_\downarrow}{\sigma_\uparrow+\sigma_\downarrow}$
is refered to as beam normal spin asymmetry. It was shown that such an 
asymmetry, to leading order in the electromagnetic coupling constant 
$\alpha_{em}\approx1/137$, is given by
\beqn
B_n\;=\;
\frac{\sum_{spins}2{\rm Im}({\cal{M}}^*_{1\gamma}{\rm Abs}{\cal{M}}_{2\gamma})}
{\sum_{spins}|{\cal{M}}_{1\gamma}|^2}.
\eeqn
\indent
The absorptive part of the two-photon exchange (TPE) graph is given by 
\beqn
{\rm Abs} {\cal{M}}_{2\gamma}&=&e^4
\int\frac{|\vec{k}_1|^2d|\vec{k}_1|d\Omega_{k_1}}{2E_1(2\pi)^3}
\bar{u}'\gamma_\nu(\keldagger_1+m)\gamma_\mu u\nn\\
&&\;\;\;\;\cdot\,\frac{1}{Q_1^2Q_2^2}{\rm Abs}M^{\mu\nu}(w^2,t,Q_1^2,Q_2^2),
\label{eq:im2gamma}
\eeqn
where $M^{\mu\nu}$ is the VVCS tensor. Furthermore, the asymmetry can be 
written as 
\beqn
B_n\;=\;\frac{e^2t}{D(s,t)}
\int\frac{|\vec{k}_1|^2d|\vec{k}_1|d\Omega_{k_1}}{2E_1(2\pi)^3}
\frac{1}{Q_1^2Q_2^2}\,{\rm Im}L_{\alpha\mu\nu}H^{\alpha\mu\nu}.
\eeqn

The leptonic tensor $L_{\alpha\mu\nu}$ is given by
\beqn
L_{\alpha\mu\nu}\;=\;
{\rm Tr}(\keldagger'+m)\gamma_\nu(\keldagger_1+m)\gamma_\mu\gamma_5\xidagger
(\keldagger+m)\gamma_\alpha,
\eeqn
while the hadronic tensor is defined as
\beqn
H^{\alpha\mu\nu}\;=\;\sum_{spins}{\rm Abs}M^{\mu\nu}
(\bar{N}'\Gamma^\alpha N)^*,
\eeqn
with $\Gamma^\alpha\,=\,G_M(-t)\gamma^\alpha-F_2(-t){P^\alpha\over M}$. 
In the following, we will use the hadronic tensor in the form introduced in 
the previous section for the most general inelastic case. The structure of 
the tensor of Eq.(\ref{eq:vvcstensor}) can be represented as 
\beqn
M^{\mu\nu}&=&A^{\mu\nu}\cdot \bar{N}'N\,+\,B^{\mu\nu}\cdot 
\bar{N}'\kdagger N\nn\\
&+&C^{\mu\nu} \cdot\bar{N}'i\gamma_5N\,+\,
D^{\mu\nu}\cdot \bar{N}'\ndagger N.
\label{eq:abcd}
\eeqn

The tensors $(A,B,C,D)^{\mu\nu}$ are defined in Appendix.
The sum over spins in the hadronic tensor can be readily performed with the 
result:
\beqn
H^{\alpha\mu\nu}&=&8MP^\alpha
\left[G_E A^{\mu\nu}\,+\,\frac{P\tilde{K}}{M}F_1 B^{\mu\nu}\right]\\
&+&8L^2G_M
\left[\left(\tilde{K}^\alpha-\frac{L\tilde{K}}{L^2}L^\alpha\right)B^{\mu\nu}
\,+\,n^\alpha D^{\mu\nu}\right]\nn.
\eeqn

Notice that the tensor $C^{\mu\nu}$ appears with $\gamma_5$, and does not 
contribute to this observable. We list here the result of the tensor 
contraction and refer the reader to Appendix for the details of the 
calculation:
\beqn
&&L_{\alpha\mu\nu}H^{\alpha\mu\nu}\;=\;64imM(n\cdot\xi)
\label{eq:bn_general}\\
&&\times
\left\{G_E\left(B_1+B_3-\frac{L^2Q_1^2Q_2^2}{\tilde{K}^2(L'^2)^2}B_{17}\right)
\right.\nn\\
&&\;+\,\frac{P\tilde{K}}{M}F_1
\left(B_2+B_4-\frac{L^2Q_1^2Q_2^2}{\tilde{K}^2(L'^2)^2}B_{18}\right)\nn\\
&&\;-\,\frac{2(P'k_1)(P\tilde{K})}{P'^2\tilde{K}^2}
\left(G_EB_1+\frac{P\tilde{K}}{M}F_1B_2\right)\nn\\
&&\;+\,\frac{Q_1^2}{n^2}
\left[PK-P\tilde{K}\frac{(q_2L) L^2}{\tilde{K}^2L'^2}\right]
\left(G_EB_9+\frac{P\tilde{K}}{M}F_1B_{10}\right)\nn\\
&&\;+\,\left.\frac{Q_2^2}{n^2}
\left[PK+P\tilde{K}\frac{(q_1L) L^2}{\tilde{K}^2L'^2}\right]
\left(G_EB_{13}+\frac{P\tilde{K}}{M}F_1B_{14}\right)\right\}\nn\\
&&+128imL^2G_M\frac{(n\xi)}{P'^2}\nn\\
&&\times\left\{B_8-(P'K)B_2+\frac{Q_1^2+Q_2^2}{16\tilde{K}^2L'^2}
(Q_1^2B_{10}+Q_2^2B_{14})\right\}
\nn\\
&&-128imL^2\frac{(n k_1)(k_1\xi)}{n^2}\nn\\
&&\times\left\{
M^3G_E\left[(1-{t\over4M^2})B_3+\frac{P\tilde{K}}{M}B_4\right]-L^2G_MB_6
\right.\nn\\
&&\;\;\;\;\;+\,\left.G_M\left[(L\tilde{K})B_8+
\frac{(P\tilde{K})L^2}{2\tilde{K}^2L'^2}(Q_1^2B_{12}+Q_2^2B_{16})\right]
\right\}\nn.
\eeqn

For forward kinematics, the virtualities of the photons are very close to 
each other such that the contributions of the terms $\sim(Q_1^2-Q_2^2)^n$ are 
suppressed by powers of $t/s$. In this case, we can use realtions between the 
amplitudes $B_i$ established in the previous sections. Denoting 
$Q_1^2=Q_2^2\equiv Q^2$, we obtain:
\beqn
&&L_{\alpha\mu\nu}H_{EVVCS}^{\alpha\mu\nu}\;=\;64imM(n\cdot\xi)
\label{eq:bn_forward}\\
&&\times
\left\{G_E\left(B_1+B_3-\frac{Q^4}{\tilde{K}^2L'^2}B_{17}\right)
\right.\nn\\
&&\;+\,\frac{P\tilde{K}}{M}F_1
\left(B_2+B_4-\frac{Q^4}{\tilde{K}^2L'^2}B_{18}\right)\nn\\
&&\;-\,\frac{2(P'k_1)(P\tilde{K})}{P'^2\tilde{K}^2}
\left(G_EB_1+\frac{P\tilde{K}}{M}F_1B_2\right)\nn\\
&&\;+\,\frac{2Q^2}{n^2}\left.
\left[PK+P\tilde{K}\frac{L^2}{\tilde{K}^2}\right]
\left(G_EB_9+\frac{P\tilde{K}}{M}F_1B_{10}\right)\right\}\nn\\
&&-128imL^2G_M(n\xi)
\left\{\frac{(P'K)}{P'^2}B_2+\frac{Q^4}{4n^2}B_{10}\right\}
\nn\\
&&-128imL^2\frac{(n k_1)(k_1\xi)}{n^2}\nn\\
&&\times\left\{
M^3G_E\left[(1-{t\over4M^2})B_3+\frac{P\tilde{K}}{M}B_4\right]
\right.\nn\\
&&\;\;\;\;\;-\,\left.G_M
\left[L^2B_6-(P\tilde{K})\frac{Q^2}{\tilde{K}^2}B_{12}\right]
\right\}\nn
\eeqn

A further simplification can be obtained in the quasi-RCS approximation. 
It amounts to considering a soft intermediate electron. This kinematical 
region is enhanced by large logarithms $\sim\ln(-t/m^2)$. This approximation 
corresponds to taking $k_1=0$ and $Q_1^2=Q_2^2=0$ in the numerator.
\beqn
&&L_{\alpha\mu\nu}H_{qRCS}^{\alpha\mu\nu}\;=\;64imM(n\cdot\xi)
\label{eq:bn_qrcs}\\
&&\times
\left\{G_E\left(B_1+B_3\right)
\;+\,\frac{P\tilde{K}}{M}F_1
\left(B_2+B_4\right)\right\}\nn.
\eeqn

\section{Calculation of $B_n$ in the forward kinematics}
In this section, we are going to apply the formalism developed in the previous 
section to a calculation with Regge kinematics. Recent calculations have shown 
that in the forward kinematics, a logarithmic enhancement takes place 
\cite{afanas_qrcs}, but the contribution of the $\ln^2(-t/m^2)$ term comes with 
photon helicity flip amplitude of RCS and its contribution is negligibly small 
\cite{afanas_erratum, ja_qrcs}. 
Vanishing of the RCS contribution at the exact quasi-RCS 
point requires that we take into account the full $k_1$ and $Q_{1,2}^2$ 
dependence 
of the hadronic tensor. The only simplification which arises due to the 
forward kinematics is based on the observation that integrals containing 
powers of $(Q_1^2-Q_2^2)$ are suppressed by powers of the (small) momentum 
transfer $t$. It can be shown, in fact, that
\beqn
&&\int\frac{|\vec{k}_1|^2d|\vec{k}_1|d\Omega_{k_1}}{2E_1(2\pi)^3}
\frac{Q_1^2-Q_2^2}{Q_1^2Q_2^2}\;=\;0,\nn\\
&&\int\frac{|\vec{k}_1|^2d|\vec{k}_1|d\Omega_{k_1}}{2E_1(2\pi)^3}
\frac{(Q_1^2-Q_2^2)^2}{Q_1^2Q_2^2}\;=\;{\cal{O}}(t/s),
\eeqn
and we refer the reader to the Appendix for details of the calculation.
Therefore, in the Regge regime, elastic VVCS should give the leading $t$ 
contribution, and we can use the hadronic tensor in the form of 
Eq.(\ref{eq:evvcstensor}) whose contribution to $B_n$ is given by 
Eq.(\ref{eq:bn_forward}). In order to calculate the asymmetry $B_n$ in these 
kinematics, we are going to relate the invariant amplitudes $B_i$ to the 
helicity amplitudes of doubly VCS. The imaginary parts of these latter are 
related to the total photoabsorption cross section by means of the optical 
theorem.
\subsection{Helicity amplitudes of VVCS with $Q_1^2=Q_2^2$}
For this calculation, we should specify the reference frame in which we work, 
since the helicity amplitudes are frame dependent. This frame dependence is 
then eliminated with use of the invariant amplitudes. For the calculation, 
the Breit frame will be used \footnote{Note that the $\gamma^*p$ c.m. frame 
cannot be used here, since in this frame it is impossible to fix the 
electrons' kinematics to the experimental one}. It is defined by 
$\vec{p}+\vec{p'}=0$.
Furthermore,
\beqn
P^\mu&=&(\sqrt{M^2-\frac{t}{4}},\vec{0}),\\
\tilde{K}^\mu&=&(\omega,0,0,q),\nn\\
L^\mu&=&(0,\frac{\Delta}{2},0,0),\nn\\
q_1&=&(\omega,\frac{\Delta}{2},0,q),\nn\\
q_2&=&(\omega,-\frac{\Delta}{2},0,q),\nn
\eeqn
where $\Delta=\sqrt{-t}>0$.
We next define the polarization vectors of the virtual photons with the 
photon helicity $\lambda_1=\pm1,0$ referring to the initial photon and 
$\lambda_2=\pm1,0$ to the final photon:
\beqn
\varepsilon_1^\mu(\lambda_1=\pm1)&=&-\frac{1}{\sqrt{2}}
\left(0,\frac{q}{q_1},i\lambda_1,-\frac{\Delta}{2q_1}\right),\nn\\
\varepsilon_1^\mu(\lambda_1=0)&=&\frac{1}{\sqrt{Q^2}}
\left(q_1,\frac{\omega}{q_1}\vec{q_1}\right),\\
\varepsilon_2^\mu(\lambda_2=\pm1)&=&-\frac{1}{\sqrt{2}}
\left(0,\frac{q}{q_1},i\lambda_2,\frac{\Delta}{2q_1}\right),\nn\\
\varepsilon_2^\mu(\lambda_2=0)&=&\frac{1}{\sqrt{Q^2}}
\left(q_1,\frac{\omega}{q_1}\vec{q_2}\right),\\
\eeqn
where $q_1$ denotes the magnitude of the photon three-vector, 
$q_1=|\vec{q_1}|=|\vec{q_2}|=\sqrt{q^2-t/4}$.
The Compton helicity amplitudes are defined as 
\beqn
T_{\lambda_2\lambda';\lambda_1\lambda}\;=\;e^2
\varepsilon_1^\mu(\lambda_1)
(\varepsilon_2^\nu(\lambda_2))^*\cdot M_{\mu\nu},
\eeqn
where we explicitly took out the electric charge $e$. We next define shorthands,
\beqn
T_{LL}&=&T_{0\lambda';0\lambda}\nn\\
T_{TT}&=&T_{\pm\lambda';\pm\lambda}\nn\\
T_{TL}&=&T_{\pm\lambda';0\lambda}\nn\\
T_{LT}&=&T_{0\lambda';\pm\lambda}.
\eeqn

With these definitions, we obtain:
\beqn
\frac{1}{e^2}T_{LL}&=&\frac{4Q^2}{t+4Q^2}\bar{N}'_{\lambda'}
\left[
\frac{q^2}{q_1^2}(B_1+\kdagger B_2)
\,-\,\frac{\omega^2}{q_1^2}(B_{17}+\kdagger B_{18})\right.\nn\\
&&\;\;\;\;\;\;\;\;\;\;\;\;\;\left.
\,-\,\frac{2\omega}{\sqrt{M^2-\frac{t}{4}}q_1^2}(B_9+\kdagger B_{10})
\right]N_\lambda,\label{eq:tll}
\eeqn
\beqn
\frac{1}{e^2}T_{TT}&=&\frac{1}{2}\bar{N}'_{\lambda'}
\left[
\frac{\omega^2}{q_1^2}\frac{t}{t+4Q^2}(B_1+\kdagger B_2)\right.\nn\\
&&\;\;\;\;\;\;\;\;\;\;
\,-\,\lambda_1\lambda_2(B_3+\kdagger B_4)\nn\\
&&\;\;\;\;\;\;\;\;\;\;
\,+\,\frac{4Q^2}{t+4Q^2}\frac{2\omega}{\sqrt{M^2-\frac{t}{4}}q_1^2}
(B_9+\kdagger B_{10})\nn\\
&&\;\;\;\;\;\;\;\;\;\;\,-\,\frac{16Q^4}{t(t+4Q^2)}
\frac{q^2}{q_1^2}(B_{17}+\kdagger B_{18})\nn\\
&&\;\;\;\;\;\;\;\;\;\;
\,-\,(\lambda_1+\lambda_2)\frac{1}{\sqrt{M^2-\frac{t}{4}}q_1}\,i\ndagger\nn\\
&&\;\;\;\;\;\;\;\;\;\;\,\times\,\left(
\frac{\omega}{\sqrt{M^2-\frac{t}{4}}q_1^2}B_6\,+\,\frac{16Q^2}{t(t+4Q^2)}B_{12}
\right)\nn\\
&&\;\;\;\;\;\;\;\;\;\;\,+\,
(\lambda_2-\lambda_1)\frac{1}{\sqrt{M^2-\frac{t}{4}}q_1}\gamma_5
\label{eq:ttt}\\
&&\;\;\,\left.\times\,
\left(
\frac{2\omega}{\sqrt{M^2-\frac{t}{4}}q_1^2}B_7
\,+\,\frac{16Q^2}{t(t+4Q^2)}B_{11}
\right)
\right]N_\lambda,\nn
\eeqn
\beqn
\frac{1}{e^2}T_{TL}&=&-\frac{\sqrt{2Q^2}\omega q}{\Delta q_1^2}
\bar{N}'_{\lambda'}
\left[
\frac{t}{t+4Q^2}(B_1+\kdagger B_2)\right.\label{eq:ttl}\\
&&\;\;\;\;\;\;\;\;\;\;
\,+\,\frac{4Q^2q^2-t\omega^2}{q^2(t+4Q^2)}
\frac{1}{\sqrt{M^2-\frac{t}{4}}\omega}(B_9+\kdagger B_{10})\nn\\
&&\;\;\;\;\;\;\;\;\;\;\;\;\;\;\left.\,+\,\frac{4Q^2}{t+4Q^2}
\frac{q^2}{q_1^2}(B_{17}+\kdagger B_{18})\right]N_\lambda,\nn\\
&-&\frac{\sqrt{2Q^2}\lambda_2}{(M^2-\frac{t}{4})\Delta qq_1}
\bar{N}'_{\lambda'}(\gamma_5B_7-i\ndagger B_6)N_\lambda\nn\\
&+&
\frac{4\sqrt{2Q^2}\omega\lambda_2\bar{N}'_{\lambda'}
(\gamma_5B_{11}-i\ndagger B_{12})
N_\lambda}{\sqrt{M^2-\frac{t}{4}}\Delta qq_1(t+4Q^2)}
,\nn
\eeqn
\beqn
\frac{1}{e^2}T_{LT}&=&\frac{\sqrt{2Q^2}\omega q}{\Delta q_1^2}
\bar{N}'_{\lambda'}
\left[
\frac{t}{t+4Q^2}(B_1+\kdagger B_2)\right.\label{eq:tlt}\\
&&\;\;\;\;\;\;\;\;
\,+\,\frac{4Q^2q^2-t\omega^2}{q^2(t+4Q^2)}
\frac{1}{\sqrt{M^2-\frac{t}{4}}\omega}(B_9+\kdagger B_{10})\nn\\
&&\;\;\;\;\;\;\;\;\;\;\;\;\;\;\left.\,+\,\frac{4Q^2}{t+4Q^2}
\frac{q^2}{q_1^2}(B_{17}+\kdagger B_{18})\right]N_\lambda,\nn\\
&-&\frac{\sqrt{2Q^2}\lambda_1}{(M^2-\frac{t}{4})\Delta qq_1}
\bar{N}'_{\lambda'}(\gamma_5B_7+i\ndagger B_6)N_\lambda\nn\\
+&&\!\!\!\!
\frac{4\sqrt{2Q^2}\omega\lambda_1}{\sqrt{M^2-\frac{t}{4}}\Delta qq_1(t+4Q^2)}
\bar{N}'_{\lambda'}
(\gamma_5B_{11}+i\ndagger B_{12})
N_\lambda,\nn
\eeqn

We next note that the amplitudes $B_7$ and $B_{11}$ do not contribute to 
the beam asymmetry. As can be seen from Eq.(\ref{eq:ttt}), the amplitudes 
$B_6$ and $B_{12}$ depend linearly on the photon helicities and therefore 
drop out of the "cross section combination" of the helicity amplitudes, 
$(T_{1\lambda';1\lambda}+T_{-1\lambda';-1\lambda})$. Instead, they 
obtain their contribution from the GDH sum rule-like combination
$(T_{1\lambda';1\lambda}-T_{-1\lambda';-1\lambda})$ which vanishes 
in the considered kinematical regime according to Pomeranchuk's theorem. 
In previous work, the contribution of the photon helicity-flip 
Compton amplitudes 
was studied for the beam asymmetry \cite{ja_qrcs, jaqrcs2}. 
In spite of double-logarithmic enhancement due to the quasi-RCS kinematics, 
their contribution to the asymmetry 
$B_n$ was found to be negligibly small for forward kinematics, and will be 
neglected in the following. \\ 
\indent
The longitudinal-transverse amplitudes are in 
general non-zero. However, it is only the combination $T_{TL}-T_{LT}$, with the 
conserved photon helicity $\lambda_1=\lambda_2$ that contributes to the 
amplitudes $B_{1,2,3,4,9,10,17,18}$ relevant for the asymmetry. As a 
microscopic calculation shows (see, for instance, \cite{brodsky}) this 
difference vanishes for diffractive kinematics. 
Thus, the only non-zero helicity combinations are
$\Sigma_{TT}\equiv{1\over2}(T_{1\lambda';1\lambda}+T_{-1\lambda';-1\lambda})$
and $T_{LL}$. Their imaginary parts are related to the transverse and 
longitudinal cross sections, respectively, which contain all the physical 
information on forward doubly virtual Compton scattering with an unpolarized 
target. 
For the remaining amplitudes $B_i$, we define shorthands, for example 
${\bf B_1}$ for $\bar{N}'_{\lambda'}(B_1+\kdagger B_2)N_\lambda$, and find 
the expressions as functions of the helicity amplitudes:
\beqn
e^2{\bf B_1}&=&\frac{\omega^2}{q_1^2}\frac{t}{t+4Q^2}\Sigma_{TT}
\,+\,\frac{q^2}{q_1^2}\frac{4Q^2}{t+4Q^2}T_{LL},\nn\\
e^2{\bf B_3}&=&-\Sigma_{TT},\nn\\
e^2{\bf B_{9}}&=&(P\tilde{K})\frac{q^2}{q_1^2}\frac{t}{t+4Q^2}
(\Sigma_{TT}-T_{LL}),\label{eq:bi_helampl}
\\
e^2\frac{4Q^2}{t}{\bf B_{17}}&=&-\frac{q^2}{q_1^2}\frac{4Q^2}{t+4Q^2}\Sigma_{TT}
\,-\,\frac{\omega^2}{q_1^2}\frac{t}{t+4Q^2}T_{LL}.\nn
\eeqn

Working out the spinors, we get in the Breit frame:
\beqn
\bar{N}'N&=&2\sqrt{M^2-{t\over4}}2\lambda\delta_{\lambda-\lambda'},\nn\\
\bar{N}'\tilde{\kdagger}N&=&2M\omega2\lambda\delta_{\lambda-\lambda'}
\,+\,q\Delta2\lambda\delta_{\lambda\lambda'}.
\eeqn

We see that in the forward limit $t=0$, only nucleon helicity-flip 
amplitudes survive.This should not mislead the reader since the reason for this 
helicity flipping is the use of Breit frame kinematics for the nucleons:
the initial and final nucleon three-momenta are opposite by definition, 
therefore unchanged spin projection onto the $x$-axis will correspond to 
opposite helicities.
Furthermore, the optical theorem relates the imaginary parts of the forward 
amplitudes to the corresponding cross sections,
\beqn
{\rm Im}{1\over2}
(T_{1-\frac{1}{2};1\frac{1}{2}}+T_{-1-\frac{1}{2};-1\frac{1}{2}})
&=&2\sqrt{M^2-t/4}\omega\sigma_T\nn\\
{\rm Im}T_{0-\frac{1}{2};0\frac{1}{2}}
&=&2\sqrt{M^2-t/4}\omega\sigma_L
\eeqn

If the Callan-Gross relation holds, the longitudinal cross section is related 
to the transvers one as $\sigma_L=\frac{Q^2}{q_1^2}\sigma_T$.
Inserting these expressions into the tensor contraction of 
Eq.(\ref{eq:bn_forward}) and noting that
Abs$M^{\mu\nu}=-2$Im$M^{\mu\nu}$, we obtain:
\beqn
&&L_{\alpha\mu\nu}H_{EVVCS}^{\alpha\mu\nu}\;=\;=-128imMG_E\omega
(n\cdot\xi)\nn\\
&&\times\,\left[
\frac{2\omega E_1}{q^2}-\frac{Q^2}{q^2}-\frac{Q^2}{q_1^2}
+\frac{Q^4}{q_1^4}\right]\frac{1}{e^2}\sigma_T\nn\\
&&\,+\,128imM^3L^2G_E\frac{(n k_1)(k_1\xi)}{n^2}\omega\frac{1}{e^2}\sigma_{T}.
\label{eq:bn_crossec}
\eeqn

It can now be seen that spurious singularities $\sim\frac{1}{t+4Q^2}$ 
appearing in the individual amplitudes in Eq.(\ref{eq:bi_helampl}) cancel 
in the observable, as they should. 

We next perform the integration over the electron phase space. We remind 
the reader that in the choosen kinematics
\beqn
n^\mu&=&(0,0,-\sqrt{M^2-{t\over4}}q{\Delta\over2},0),\nn\\
\xi^\mu&=&(0,0,1,0),
\eeqn
therefore
\beqn
&&\int\frac{|\vec{k}_1|^2d|\vec{k}_1|d\Omega_{k_1}}{2E_1(2\pi)^3}
\frac{1}{Q_1^2Q_2^2}\,{\rm Im}L_{\alpha\mu\nu}H^{\alpha\mu\nu}\\
&&=\,-128mMG_E\sqrt{M^2-{t\over4}}{\Delta\over2}\frac{1}{e^2}\sigma_T\nn\\
&&\left\{
2E I_1^0\,-\,2J_2^{00}\,-\,2I_2\,+\,I_3
\,-\,\frac{2M^2}{M^2-{t\over4}}J_{25}\right\},\nn
\label{eq:LH_integrals}
\eeqn
where we have used the notation:
\beqn
I_1^\mu&=&\int\frac{|\vec{k}_1|^2d|\vec{k}_1|d\Omega_{k_1}}{2E_1(2\pi)^3}
\frac{k_1^\mu}{Q_1^2Q_2^2}\nn\\
I_2&=&\int\frac{|\vec{k}_1|^2d|\vec{k}_1|d\Omega_{k_1}}{2E_1(2\pi)^3}
\frac{1}{2}\left[\frac{1}{Q_1^2}+\frac{1}{Q_2^2}\right]\nn\\
I_3&=&\int\frac{|\vec{k}_1|^2d|\vec{k}_1|d\Omega_{k_1}}{2E_1(2\pi)^3}
\frac{(Q_1^2+Q_2^2)^2}{4Q_1^2Q_2^2}
\frac{1}{(\vec{k}-\vec{k}_1)^2}\nn\\
J_2^{\mu\nu}&=&\int\frac{|\vec{k}_1|^2d|\vec{k}_1|d\Omega_{k_1}}{2E_1(2\pi)^3}
\frac{k_1^\mu k_1^\nu}{Q_1^2Q_2^2}.
\eeqn

Furthermore, the coefficient $J_{25}$ originates from decomposing the 
tensor integral according to \cite{passarino_veltman},
\beqn
J_2^{\mu\nu}&=&J_{21}P^\mu P^\nu\,+\,J_{22}K^\mu K^\nu
\,+\,J_{23}(P^\mu K^\nu+K^\mu P^\nu)\nn\\
&+&J_{24}L^\mu L^\nu\,+\,J_{25}g^{\mu\nu}.
\eeqn

The integrals $I_1^0$ and $I_2$ entering Eq.(\ref{eq:LH_integrals}) were 
calculated in the previous work \cite{ja_qrcs}. We obtain for the 
first three integrals
\beqn
I_1^0&=&\frac{E}{-4\pi^2t}
\left\{{{E_{thr}}\over E}\ln{\sqrt{-t}\over m}{{E_{thr}}\over E}\right.\\
&&\;\;\;\;\;\;\;\;\;\;\;\;\left.\,+\,
\left(1-{{E_{thr}}\over E}\right)\ln\left(1-{{E_{thr}}\over E}\right)
\right\},\nn\\
I_2&=&\frac{1}{4\pi^2}{{E_{thr}}\over E}\ln{{2{E_{thr}}\over m}}\nn\\
&+&\frac{1}{4\pi^2}
\left(1-{{E_{thr}}\over E}\right)\ln\left(1-{{E_{thr}}\over E}\right),
\nn\\
I_3&=&\frac{1}{8\pi^2}\left(1+\frac{E_{thr}}{E}\right)
\ln\left(1+\frac{E_{thr}}{E}\right)\nn\\
&+&\frac{1}{8\pi^2}\left(1-\frac{E_{thr}}{E}\right)
\ln\left(1-\frac{E_{thr}}{E}\right),
\eeqn
and for the tensor integrals:
\beqn
J_2^{00}&=&\frac{E_{thr}^2}{-8\pi^2t}
\left\{
\ln\left(\frac{Q}{m}\frac{E_{thr}}{E}\right)\,+\,\frac{E}{E_{thr}}
\right\}\\
&+&\frac{E^2}{-8\pi^2t}\left(1-\frac{E_{thr}^2}{E^2}\right)
\ln\left(1-\frac{E_{thr}}{E}\right),\nn\\
J_{25}&=&\frac{1}{32\pi^2}\frac{E_{thr}^2}{E^2+t/4}
\left[\ln\frac{\sqrt{-t}}{2E}+\frac{E}{E_{thr}}+\frac{1}{2}\right.\nn\\
&&+\left.\left(\frac{E^2}{E_{thr}^2}-1\right)\ln\left(1-\frac{E_{thr}}{E}\right)
\right]\,-\,\frac{E_{thr}^2}{32\pi^2(pk)}.\nn
\eeqn

In the above formulas, $E_{thr}$ denotes the upper limit in the integral over 
the electron energy, which corresponds to threshold production of the pion,
$E_{thr}=\frac{s-(M+m_\pi^2)^2}{2\sqrt{M^2-t/4}}$ in the Breit frame.
We refer the reader to the Appendix for details of the calculation.
We can finally write down the final result for normal beam spin asymmetry:
\beqn
B_n&=&-\frac{m\sqrt{-t}\sigma_T}{4\pi^2(1+\tau)^2}
\frac{G_E}{\tau G_M^2+\varepsilon G_E^2}\label{eq:Bn_fullresult}\\
&\cdot&
\left\{
\ln\left(\frac{\sqrt{-t}}{m}\right)
-1+\frac{t}{E^2}\ln\frac{E^2}{m^2}\right.\nn\\
&&\left.\;\;\;\;+\frac{t}{4E^2}
\left[\ln\frac{\sqrt{-t}}{2E}+1+\frac{M^2}{2s}\right]
\right\}.\nn
\eeqn

\section{Results}
The model used in the previous section for the calculation of the $B_n$ is 
based on the optical theorem, Callan-Cross relation between the longitudinal 
and transverse partial cross sections, and the neglection of the mismatch 
between the virtualities of the virtual photons inside the loop. This latter 
assumption restricts the applicability of the above formula to low values of 
the momentum transfer. To estimate the error introduced by neglecting the terms 
$\sim \frac{(Q_1^2-Q_2^2)^2}{Q_1^2Q_2^2}$ under the integral over the electron 
phase space, we note that the integrals entering the final result of 
Eq. (\ref{eq:Bn_fullresult}) have the order $\sim t^{-1}$ ($I_1^0,J_{2}^{00}$) and 
$t^0$ ($I_2,I_3,J_{25}$). As it was noticed before, terms 
$\sim \frac{(Q_1^2-Q_2^2)^2}{Q_1^2Q_2^2}$ contribute to the order $t^1$. 
Therefore, they contribute at relative order of $t^2$. Thus, for low values of 
$t$, our full result of 
Eq. (\ref{eq:Bn_fullresult}) is consistent with the approximation used. 

We next study the leading $t$-dependence of the beam normal spin asymmetry. It 
is given by
\beqn
B_n=-\frac{m\sqrt{-t}\sigma_T}{4\pi^2}
\frac{G_E}{\tau G_M^2+\varepsilon G_E^2}
\left[\ln\left(\frac{\sqrt{-t}}{m}\right)-1\right]
\label{eq:Bn_smallt}
\eeqn

This result is in exact agreement with the result of \cite{afanas_erratum}.
The approach of \cite{afanas_erratum} consisted of taking only one amplitude 
for elastic doubly VCS, the one which survives in the exact forward RCS. It 
therefore amounts to neglecting not only the additional $t$ dependence of 
the Compton amplitude, but also the dependence on the photon virtualities. 
The present calculation, on the contrary, uses the complete set of invariant 
amplitudes for the case of equal virtualities of the incoming and outgoing 
photons, and does not apriori set any kinematical variable to zero. 
The reason for this agreement is that every factor $\sim Q_{1,2}^2$ in the 
numerator actually leads to an extra suppression in $t$, as can be seen, 
if comparing the integrals $I_2$ to $I_0$ and $B_1^\mu$ to $I_1^\mu$, 
respectively. The integrals are given in the Appendix.

\begin{figure}[h]
{\includegraphics[height=9cm]{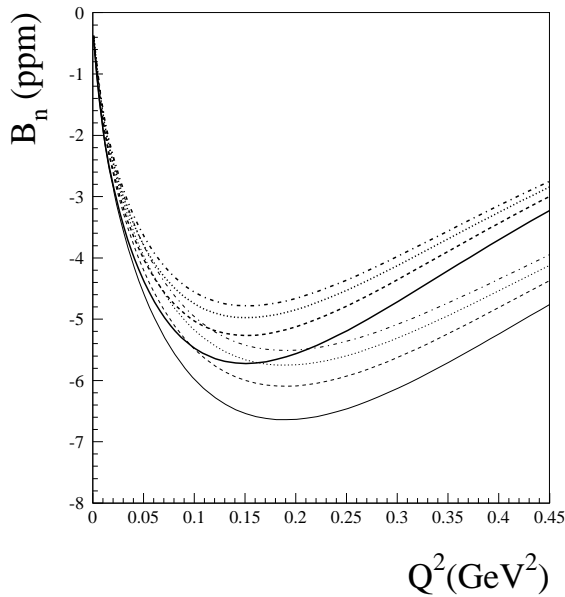}}
\caption{The results for $B_n$ as function of the momentum transfer $Q^2$ are 
shown for four values of the $lab$ beam energy: 3 GeV (solid lines), 6 GeV 
(dashed lines), 12 GeV (dotted lines), and 45 GeV (dash-dotted lines). 
The thick lines correspond to the full $t$-dependence of 
Eq.(\ref{eq:Bn_fullresult}), while the thin lines correspond to the leading 
$t$ behaviour of Eq.(\ref{eq:Bn_smallt}).}
\label{fig:Bn}
\end{figure}

If using as input the exact forward value of the imaginary part of the Compton 
amplitude, to go to finite values of $t$, one needs 
a phenomenological information in order to consistently describe the 
off-forward 
Compton scattering. We use the same approach as \cite{afanas_erratum}. It is 
based on the phenomenological exponential fit of the $t$-dependence of the 
Compton 
differential cross section for values of $t$ from 0 through -0.8 GeV$^2$.
Since $\frac{d\sigma}{dt}\sim\sigma_T^2$, one gets 
\beqn 
\sigma_T(t)\,=\,\sigma_T(0)e^{Bt/2},
\eeqn
with $B\approx5.2-8.6$ GeV$^{-2}$. 
For the energy dependence of $\sigma_T$, we use the phenomenological fit of 
Ref. \cite{sigma_tot}. We present our results in Fig. \ref{fig:Bn}.
As seen from the figure, the more accurate account on the $t$ dependence leads 
to faster decreasing of the asymmetry with increasing $Q^2$. We predict 
therefore lower negative values for $B_n$ than the authors of 
\cite{afanas_erratum}. 
\begin{figure}[h]
{\includegraphics[height=9cm]{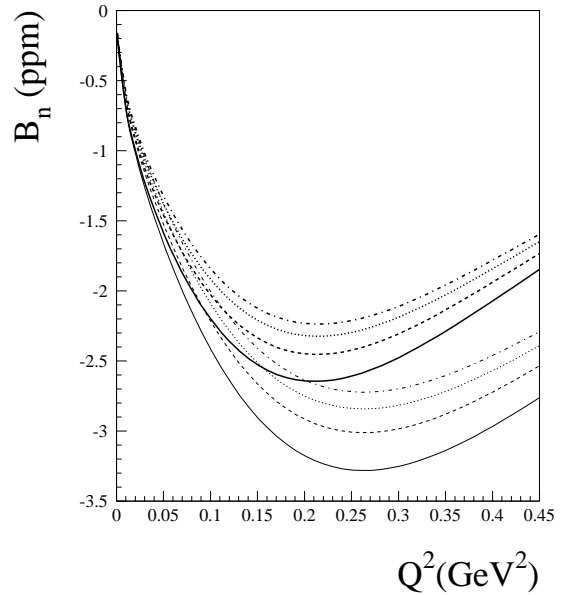}}
\caption{The results for $B_n$ as function of the momentum transfer $Q^2$ 
for the neutron target. Notation as in Fig. \ref{fig:Bn}.}
\label{fig:Bn_neutron}
\end{figure}

In Fig. \ref{fig:Bn_neutron}, results for $B_n$ on the neutron target are 
shown. 
The asymmetry is only about twice smaller for the neutron 
than for the proton, which is due to the overall factor of 
$\frac{G_E}{\tau G_M^2+\varepsilon G_E^2}$. Since also the cross section 
in the denominator is 
much smaller for the neutron target at low $Q^2$ than for the proton, 
it means that a measurement of the cross section difference in the numerator of 
the $B_n$ would be a very challenging task for the neutron target.

\section{Summary}
In summary, we constructed the invariant basis for doubly virtual Compton 
scattering (VVCS). In the general case, VVCS is described by means of 18 
independent invariant amplitudes which are functions of the invariants 
$P\tilde{K},t,q_1^2,q_2^2$. The proposed form of the VVCS tensor does not 
exist in the literature. This form is based on Lorentz and gauge invariance 
and consists of constructing 9 orthogonal basis tensors, which are useful for 
practical calculations. We furthermore investigated the behaviour of the 
VVCS amplitudes under photon and nucleon crossing. Basing on this crossing 
behaviour, it is possible to obtain the correct form of the VVCS basis in 
the case of elastic VVCS (i.e., $q_1^2=q_2^2$) where only 12 amplitudes 
survive. The special case of elastic VVCS is realized in the case of the 
calculation of the beam normal spin asymmetry $B_n$ with forward kinematics. 
We provide the calculation for this observable with these kinematics. Since 
in the 
exact forward limit this observable vanishes, one has to use the full amplitude 
for this calculation. We project the Compton helicity amplitudes onto the 
invariant basis and use as input the optical theorem. It relates the imaginary 
part of the helicity amplitudes 
${1\over2}(T_{1{1\over2},1{1\over2}}+T_{-1{1\over2},-1{1\over2}})$ and 
$T_{0{1\over2},0{1\over2}}$ to the transverse $\sigma_T$ and the longitudinal 
$\sigma_L$ cross sections, respectively. The longitudinal cross section is then 
related to the transverse one by means of the Callan-Cross relation. The 
resulting 
asymmetry is of the order of $\approx -5$ ppm and is practically independent of 
the electron energy. 

The presented formalism may be furthermore extended to 
the two photon exchange contributions to other observables. 
This work was supported by the US Department of Energy 
contract DOE-FG02-05ER 41361.

\section{Appendix}
\subsection{Projection technique}
In a real calculation, one needs to calculate the single amplitudes 
contributions from, for instance, a Feynman diagram. For this, one can make 
use of the fact that all the basis tensors are orthogonal. For example, 
\beqn
\frac{P'^\mu P'^\nu}{P'^2}W_{\mu\nu}
\,=\,\bar{N}'\left[B_1+\tilde{\kdagger}B_2\right]N,
\eeqn
and so on. We find following useful relations:
\beqn
P'^2&=&P^2-\frac{P\tilde{K}}{\tilde{K}^2}\frac{L^2}{L'^2}\nn\\
L'^2&=&L^2-\frac{(L\tilde{K})^2}{\tilde{K}^2}\nn\\
n^2&=&-P'^2\tilde{K}^2L'^2\nn\\
\tilde{K}'^2&=&-\frac{\tilde{K}^2}{L'^2}Q_1^2\nn\\
\tilde{K}''^2&=&-\frac{\tilde{K}^2}{L'^2}Q_2^2\nn\\
\tilde{K}'\cdot\tilde{K}''&=&\frac{\tilde{K}^2}{L'^2}
\left(L^2-\tilde{K}^2\right),
\eeqn
and furthermore,
\beqn
P'\cdot\tilde{K}&=&P'\cdot L'=P'\cdot L=P'\cdot\tilde{K}'=P'\cdot\tilde{K}''
=P'\cdot n\nn\\
&=&n\cdot\tilde{K}=n\cdot L'=n\cdot L=n\cdot\tilde{K}'=n\cdot\tilde{K}''
=0;\nn\\
q_1\cdot L'&=&-q_2\cdot L'=L\cdot L'=L'^2.
\eeqn

\subsection{Leptonic and hadronic tensors}
The hadronic tensors quoted in Eq.(\ref{eq:abcd}) are given by 
\beqn
A^{\mu\nu}&=&\frac{P'^\mu P'^\nu}{P'^2}B_1\;+\;
\frac{n^\mu n^\nu}{n^2}B_3\;+\frac{\tilde{K}'^\mu P'^\nu}{P'^2\tilde{K}^2}
B_9\nn\\
&+&\frac{P'^\mu\tilde{K}''^\nu}{P'^2\tilde{K}^2}B_{13}
\;+\frac{\tilde{K}'^\mu\tilde{K}''^\nu}{\tilde{K}^2}
B_{17}
\eeqn
\beqn
B^{\mu\nu}&=&\frac{P'^\mu P'^\nu}{P'^2}B_2\;+\;
\frac{n^\mu n^\nu}{n^2}B_4\;+\frac{\tilde{K}'^\mu P'^\nu}{P'^2\tilde{K}^2}
B_{10}\nn\\
&+&\frac{P'^\mu\tilde{K}''^\nu}{P'^2\tilde{K}^2}B_{14}
\;+\frac{\tilde{K}'^\mu\tilde{K}''^\nu}{\tilde{K}^2}
B_{18}
\eeqn
\beqn
C^{\mu\nu}&=&\frac{P'^\mu n^\nu\,+\,n^\mu P'^\nu}{P'^2n^2}B_5
\,+\,\frac{P'^\mu n^\nu\,-\,n^\mu P'^\nu}{P'^2n^2}B_7\nn\\
&+&\frac{\tilde{K}'^\mu n^\nu}{n^2\tilde{K}^2}B_{11}
\,+\,\frac{\tilde{K}''^\nu n^\mu}{n^2\tilde{K}^2}B_{15}
\eeqn
\beqn
D^{\mu\nu}&=&\frac{P'^\mu n^\nu\,+\,n^\mu P'^\nu}{P'^2n^2}B_6
\,+\,\frac{P'^\mu n^\nu\,-\,n^\mu P'^\nu}{P'^2n^2}B_8\nn\\
&+&\frac{\tilde{K}'^\mu n^\nu}{n^2\tilde{K}^2}B_{12}
\,+\,\frac{\tilde{K}''^\nu n^\mu}{n^2\tilde{K}^2}B_{16}
\eeqn

The leptonic tensor 
\beqn
L_{\alpha\mu\nu}\;=\;
{\rm Tr}(\keldagger'+m)\gamma_\nu(\keldagger_1+m)\gamma_\mu\gamma_5\xidagger
(\keldagger+m)\gamma_\alpha
\eeqn
can be conveniently decomposed into two parts which are symmetric and 
antisymmetric in the pair of indices $(\mu\nu)$:
\beqn
L_{\alpha\mu\nu}^{SYM}&=&-8im
\left\{
g_{\mu\nu}\varepsilon_{\kappa\rho\tau\alpha}L^\kappa\tilde{K}^\rho\xi^\tau
\right.\\
&+&\left.(k_1)_\mu\varepsilon_{\kappa\nu\tau\alpha}L^\kappa\xi^\tau
\;+\;(k_1)_\nu\varepsilon_{\kappa\mu\tau\alpha}L^\kappa\xi^\tau,
\right\}\nn
\eeqn
\beqn
L_{\alpha\mu\nu}^{ASYM}&=&-8im
\left\{
\xi_\nu\varepsilon_{\kappa\mu\tau\alpha}L^\kappa K^\tau
\,-\,\xi_\mu\varepsilon_{\kappa\nu\tau\alpha}L^\kappa K^\tau
\right.\nn\\
&+&L_\alpha\varepsilon_{\nu\mu\tau\rho}\xi^\tau L^\rho
\,-\,L^2\varepsilon_{\nu\mu\tau\alpha}\xi^\tau \nn\\
&-&\left.K_\alpha\varepsilon_{\nu\mu\tau\rho}\xi^\tau \tilde{K}^\rho
\,+\,\xi_\alpha\varepsilon_{\nu\mu\tau\rho}k_1^\tau L^\rho
\right\}.
\eeqn
\subsection{Tensor contractions}
\beqn
&&\left[L^2\tilde{K}^\alpha-(L\tilde{K})L^\alpha\right]B^{\mu\nu}\cdot 
L_{\alpha\mu\nu}\,=\,-8imL^2(n\xi)\nn\\
&&\times\,\left\{\frac{2(P'k_1)}{P'^2}B_2
\,+\frac{Q_1^2+Q_2^2}{4n^2}\left[Q_1^2B_{10}+Q_2^2B_{14}\right]\right\}\nn\\
&&+\,16im(L^2)^2\frac{(nk_1)(\tilde{K}\xi)}{n^2}(P\tilde{K})B_4.
\eeqn
\beqn
&&n^\alpha D^{\mu\nu}\cdot L_{\alpha\mu\nu}
\,=\,16im\frac{(nk_1)(\tilde{K}\xi)}{n^2}\\
&&\times\,\left\{-L^2B_6\,+\,(L\tilde{K})B_8
\,+\,\frac{(P\tilde{K})L^2}{\tilde{K}^2L'^2}
\left[Q_1^2B_{12}+Q_2^2B_{16}\right]\right\}.\nn
\eeqn

\beqn
&&P^\alpha A^{\mu\nu}\cdot L_{\alpha\mu\nu}
\,=\,16im\frac{(nk_1)(\tilde{K}\xi)}{n^2}P^2L^2B_3\nn\\
&&\,+\,8im(n\xi)
\left\{
B_1\,+\,B_3\,+\,\frac{2(P'k_1)(P\tilde{K})}{n^2}B_1\right.\nn\\
&&\,+\,\left[\frac{(\tilde{K}'\tilde{K}'')}{\tilde{K}^2}+\frac{2K^2}{L'^2}
+\frac{(q_1L)(\tilde{K}''k_1)-(q_2L)(\tilde{K}'k_1)}{\tilde{K}^2L'^2}
\right] B_{17}\nn\\
&&\,+\,
\left[\frac{(P\tilde{K})L^2}{n^2}
\left(\frac{(\tilde{K}'k_1)}{\tilde{K}^2}-1\right)\right.\nn\\
&&\;\;\;\;\;\;\left.
-\frac{(PK)(q_1\tilde{K})+(P'k_1)(q_1L)}{n^2}\right]B_{10}\nn\\
&&\,+\,
\left[\frac{(P\tilde{K})L^2}{n^2}
\left(\frac{(\tilde{K}''k_1)}{\tilde{K}^2}-1\right)\right.\nn\\
&&\;\;\;\;\;\;\left.\left.
-\frac{(PK)(q_2\tilde{K})-(P'k_1)(q_2L)}{n^2}\right]B_{14}\right\}\nn\\
&&\,=\,16im\frac{(nk_1)(\tilde{K}\xi)}{n^2}P^2L^2B_3\\
&&\,+\,8im(n\xi)
\left\{
B_1\,+\,B_3\,+\,\frac{2(P'k_1)(P\tilde{K})}{n^2}B_1\right.\nn\\
&&\;\;\;\;\;\;\;\;\;\;\;\;\;\;\;\;\;\;
\,-\,\frac{L^2Q_1^2Q_2^2}{\tilde{K}^2(L'^2)^2}B_{17}\nn\\
&&\;\;\;\;\;\;\;\;\;\;\;\;\;\;\;\;\;\;\,+\,\frac{Q_1^2}{n^2}
\left[(PK)-(P\tilde{K})\frac{(q_2L)L^2}{\tilde{K}^2L'^2}\right]B_{9}\nn\\
&&\;\;\;\;\;\;\;\;\;\;\;\;\;\;\;\;\;\;\,+\,\left.\frac{Q_2^2}{n^2}
\left[(PK)+(P\tilde{K})\frac{(q_1L)L^2}{\tilde{K}^2L'^2}\right]B_{13}
\right\}\nn
\eeqn

\subsection{Integrals over the electron phase space}
The integrals appearing in the final result are:
\beqn
I_0&=&\int\frac{|\vec{k}_1|^2d|\vec{k}_1|d\Omega_{k_1}}{2E_1(2\pi)^3}
\frac{1}{Q_1^2Q_2^2}\nn\\
I_1^\mu&=&\int\frac{|\vec{k}_1|^2d|\vec{k}_1|d\Omega_{k_1}}{2E_1(2\pi)^3}
\frac{k_1^\mu}{Q_1^2Q_2^2}\nn\\
I_2&=&\int\frac{|\vec{k}_1|^2d|\vec{k}_1|d\Omega_{k_1}}{2E_1(2\pi)^3}
\frac{1}{2}\left[\frac{1}{Q_1^2}+\frac{1}{Q_2^2}\right]\nn\\
I_3&=&\int\frac{|\vec{k}_1|^2d|\vec{k}_1|d\Omega_{k_1}}{2E_1(2\pi)^3}
\frac{(Q_1^2+Q_2^2)^2}{4Q_1^2Q_2^2}
\frac{1}{(\vec{k}-\vec{k}_1)^2}\nn\\
J_2^{\mu\nu}&=&\int\frac{|\vec{k}_1|^2d|\vec{k}_1|d\Omega_{k_1}}{2E_1(2\pi)^3}
\frac{k_1^\mu k_1^\nu}{Q_1^2Q_2^2}.
\eeqn
\indent
Furthermore, the terms $\sim\frac{(Q_1^2-Q_2^2)^2}{Q_1^2Q_2^2}$, 
which we have consistently neglected throughout the calculation, lead to 
the integral 
\beqn
I_4\,=\,\int\frac{|\vec{k}_1|^2d|\vec{k}_1|d\Omega_{k_1}}{2E_1(2\pi)^3}
\frac{(Q_1^2-Q_2^2)^2}{Q_1^2Q_2^2}=16L_\mu L_\nu J_2^{\mu\nu}.
\eeqn

The first integral was already calculated \cite{afanas_qrcs, ja_qrcs}. 
We next turn to the vector integral
\beqn
I_1^\mu&=&\int_0^{k_{thr}}\frac{k_1^2dk_1}{2E_1(2\pi)^3}
\int d\Omega_{k_1}\frac{k_1^\mu}{(k-k_1)^2(k'-k_1)^2}\nn\\
&=&I_{1P}P^\mu\,+\,I_{1K}K^\mu\,.
\eeqn
\indent
It cannot depend on $q^\mu$ due to the symmetry of $I_1$ under 
interchanging $k$ and $k'$. To determine these two coefficients we have a 
system of equations,
\beqn
K^0I_{1K}\,+\,P^0I_{1P}&=&I_1^0\;=\;
{1\over16\pi^3}\int\frac{d^3\vec{k}_1}{Q_1^2Q_2^2}\\
-tI_{1K}\,+\,4PKI_{1P}&=&4K_\mu I_1^\mu\nn\\
&=&
\int\frac{d^3\vec{k}_1}{(2\pi)^3E_1Q_1^2}
\,\equiv\,I_2.\nn
\eeqn
Using the same approach as for $I_0$ (see the Appendix of \cite{ja_qrcs}), 
we obtain for $I_1^0$:
\beqn
I_1^0&=&\frac{E}{-8\pi^2t}\int_{m\over E}^{{E_{thr}}\over E}
\frac{zdz}{\sqrt{z^2-\frac{m^2}{E^2}-\frac{4m^2}{t}(1-z)^2}}\\
&&\;\;\;\;\;\;\;\;\cdot
\ln\frac{\sqrt{z^2-\frac{m^2}{E^2}-\frac{4m^2}{t}(1-z)^2}+\sqrt{z^2-\frac{m^2}{E^2}}}
{\sqrt{z^2-\frac{m^2}{E^2}-\frac{4m^2}{t}(1-z)^2}-\sqrt{z^2-\frac{m^2}{E^2}}}
\nn\\
&=&\frac{E}{-4\pi^2t}
\left\{{{E_{thr}}\over E}\ln{\sqrt{-t}\over m}{{E_{thr}}\over E}\right.\\
&&\;\;\;\;\;\;\;\;\;\;\;\;\left.\,+\,
\left(1-{{E_{thr}}\over E}\right)\ln\left(1-{{E_{thr}}\over E}\right)
\right\}.
\eeqn
\indent
Finally, we consider the integral $I_2$ where we perform the angular 
integration,
\beqn
I_2&=&{1\over(2\pi)^3}\int\frac{d^3\vec{k}_1}{E_1Q_1^2}\\
&=&
\frac{1}{8\pi^2}\int_{m\over E}^{{E_{thr}}\over E}dz
\ln\frac{z-\frac{m^2}{E^2}+\sqrt{1-\frac{m^2}{E^2}}\sqrt{z^2-\frac{m^2}{E^2}}}
{z-\frac{m^2}{E^2}-\sqrt{1-\frac{m^2}{E^2}}\sqrt{z^2-\frac{m^2}{E^2}}}.\nn
\eeqn
\indent
After integrating by parts and changing variables,  
$z={m\over E}\cosh y$ and $y=\ln t$, we arrive at
\beqn
I_2&=&\frac{1}{4\pi^2}{{E_{thr}}\over E}\ln{{2{E_{thr}}\over m}}\\
&+&\frac{1}{4\pi^2}
\left(1-{{E_{thr}}\over E}\right)\ln\left(1-{{E_{thr}}\over E}\right).
\nn
\eeqn
\indent
In these last two integrals, it is important to keep ${{E_{thr}}\over E}$ 
unequal to 1 until the end to ensure the convergence of the integral. 
Solving the system of linear equations for the coefficients, we obtain:
\beqn
I_{1P}&=&\frac{1}{8\pi^2}
\frac{s-M^2}{M^4-su}{{E_{thr}}\over E}\ln{2E\over Q}\nn\\
I_{1K}&=&\frac{1}{Q^2}I_2-\frac{4PK}{Q^2}I_{1P}.
\eeqn
\indent
We next consider the integral $I_3$ and rewrite it as 
\beqn
I_3&=&\int\frac{|\vec{k}_1|^2d|\vec{k}_1|d\Omega_{k_1}}{2E_1(2\pi)^3}
\left[1+\frac{(Q_1^2-Q_2^2)^2}{4Q_1^2Q_2^2}\right]
\frac{1}{(\vec{k}-\vec{k}_1)^2}.\nn\\
&&
\eeqn

We note that the second term in the square brackets $\sim(Q_1^2-Q_2^2)^2$ 
is of the order $t/s$. For the moment, we keep this subleading term, and 
calculate the leading in $t$ contribution:
\beqn
I_3&=&\int\frac{|\vec{k}_1|^2d|\vec{k}_1|d\Omega_{k_1}}{2E_1(2\pi)^3}
\frac{1}{(\vec{k}-\vec{k}_1)^2}\nn\\
&=&\frac{1}{8\pi^2}\int_0^{k_{M}}dz\ln\frac{1+z}{1-z}+O(t/s)\nn\\
&=&\frac{1}{8\pi^2}\left(1+\frac{E_{thr}}{E}\right)
\ln\left(1+\frac{E_{thr}}{E}\right)\nn\\
&+&\frac{1}{8\pi^2}\left(1-\frac{E_{thr}}{E}\right)
\ln\left(1-\frac{E_{thr}}{E}\right)+O(t/s).
\eeqn

As we see, the leading term contributes to $B_n$ to relative order $t^1$. 
This means that the second term should be omitted to be consistent with the 
approximation adopted.

The calculation of the tensor integral is more involved. We decompose this 
tensor into tensorial structures constructed out of external momenta
\cite{passarino_veltman},
\beqn
J_2^{\mu\nu}&=&\int\frac{|\vec{k}_1|^2d|\vec{k}_1|d\Omega_{k_1}}{2E_1(2\pi)^3}
\frac{k_1^\mu k_1^\nu}{Q_1^2Q_2^2}\nn\\
&=&J_{21}P^\mu P^{\nu}+J_{22}K^\mu K^{\nu}
+J_{23}(P^\mu K^{\nu}+P^\nu K^{\mu})\nn\\
&+&J_{24}L^\mu L^{\nu}+J_{25}g^{\mu\nu}.
\eeqn

The final result contains 
In order to calculate the coefficients $J_{2j}$, we contract the tensor 
integral with $g^{\mu\nu}$ and external momenta:
\beqn
{J_2}^\mu_\mu&=&P^2J_{21}+K^2J_{22}+2PKJ_{23}+L^2J_{24}+4J_{25}\nn\\
&=&m^2\int\frac{|\vec{k}_1|^2d|\vec{k}_1|d\Omega_{k_1}}{2E_1(2\pi)^3}
\frac{1}{Q_1^2Q_2^2}\approx0
\eeqn
\beqn
2k_\mu2k'_\nu J_2^{\mu\nu}&=&
(2PK)^2J_{21}+4(K^2)^2J_{22}\nn\\
&+&8K^2PKJ_{23}-4(L^2)^2J_{24}+8K^2J_{25}\nn\\
&=&\int\frac{|\vec{k}_1|^2d|\vec{k}_1|d\Omega_{k_1}}{2E_1(2\pi)^3}
\equiv C_0 = \frac{E_{thr}^2}{8\pi^2}
\eeqn
\beqn
2k'_\nu J_2^{\mu\nu}&=&
2PKJ_{21}P^\mu+2K^2J_{22}K^\mu\nn\\
&+&2\left[PK K^\mu+K^2P^\mu\right]J_{23}\nn\\
&-&2L^2J_{24}L^\mu+2J_{25}(K-L)^\mu\nn\\ 
&=&\int\frac{|\vec{k}_1|^2d|\vec{k}_1|d\Omega_{k_1}}{2E_1(2\pi)^3}
\frac{k_1^\mu}{Q_1^2}\equiv B_1^{\mu}.
\label{eq:j2-b1}
\eeqn

Rewriting the vector integral $B_1$ as
\beqn
B_1^{\mu}&=&
\int\frac{|\vec{k}_1|^2d|\vec{k}_1|d\Omega_{k_1}}{2E_1(2\pi)^3}
\frac{k_1^\mu}{Q_1^2}=B_{11}p^\mu+B_{12}k^\mu\nn\\
&=&B_{11}P^\mu+B_{12}K^\mu+(B_{12}-B_{11})L^\mu,
\eeqn
we find following relations between $J_{2i}$ and $B_i$:
\beqn
2PK\,J_{21}\,+\,2K^2J_{23}&=&B_{11}\nn\\
2K^2J_{22}\,+\,2PK\,J_{23}\,+\,2J_{25}&=&B_{12}\nn\\
2L^2J_{24}\,+\,2J_{25}&=&B_{11}-B_{12}.
\eeqn 

To proceed, we calculate the appearing vector integral. 
The coefficients can be found as following:
\beqn
2k_\mu B_1^\mu&=&2pkB_{11}=C_0,
\eeqn
while the $0$-component $B_1^0=p^0B_{11}+k^0B_{12}$ can be calculated 
directly: 
\beqn
B_1^0&=&
\int\frac{|\vec{k}_1|^2d|\vec{k}_1|d\Omega_{k_1}}{16\pi^3}
\frac{1}{Q_1^2}\nn\\
&=& 
\frac{E}{16\pi^2}\int_{m\over E}^{{E_{thr}}\over E}
z dz \ln\frac{z+\sqrt{z^2-\frac{m^2}{E^2}}}{z-\sqrt{z^2-\frac{m^2}{E^2}}}
\nn\\
&=&\frac{E_{thr}^2}{16\pi^2E}\left[\ln\frac{2E_{thr}}{m}-{1\over2}\right].
\eeqn

The fifth independant equation for five unknown coefficients $J_{2j}$ is 
obtained by the calculation of the $J_2^{00}$ component of the tensor integral. 
The calculation follows the same steps as that for $I_1^0$ given in the 
Appendix of \cite{ja_qrcs}. We obtain:
\beqn
J_2^{00}&=&
\int\frac{|\vec{k}_1|^2d|\vec{k}_1|d\Omega_{k_1}}{16\pi^3}E_1
\frac{1}{Q_1^2Q_2^2}\\
&=& 
\frac{E^2}{-8\pi^2t}\int_{0}^{{k_{thr}}\over k}
\frac{z^2 dz}{\sqrt{z^2-\frac{4m^2}{t}(1-z)^2}}\nn\\
&&\;\;\;\;\;\;\;\;
\ln\frac{z+\sqrt{z^2-\frac{4m^2}{t}(1-z)^2}}{\sqrt{z^2-\frac{4m^2}{t}(1-z)^2}-z}
.\nn
\eeqn

Due to two powers of $z=k_1/k$ in the numerator, the contribution of small 
values of the intermediate electron momenta is negligible, and 
the square root can be Taylor-expanded since $m^2/|t|<<1$. This leads to 
\beqn
J_2^{00}&=&\frac{E^2}{-8\pi^2t}\int_{0}^{{E_{thr}}\over E}
zdz\ln\left(\frac{-t}{m^2}\frac{z^2}{(1-z)^2}\right)\\
&=&\frac{E_{thr}^2}{-8\pi^2t}
\left[\ln\frac{\sqrt{-t}E_{thr}}{mE}+\frac{E}{E_{thr}}\right.\nn\\
&&\;\;\;\;\left.+\frac{E^2}{E_{thr}^2}\left(1-\frac{E_{thr}^2}{E^2}\right)
\ln\left(1-\frac{E_{thr}}{E}\right)\right]\nn
\eeqn

Solving the system of linear equations for the coefficients $J_{2j}$, we find 
\beqn
J_{25}&=&\frac{P^2}{P'^2}\left(J_2^{00}-\frac{E}{2K^2}B_1^0\right)
-\frac{1}{4(pk)}C_0\\
&=&\frac{1}{32\pi^2}\frac{E_{thr}^2}{E^2+t/4}
\left[\ln\frac{\sqrt{-t}}{2E}+\frac{E}{E_{thr}}+\frac{1}{2}\right.\nn\\
&&+\left.\left(\frac{E^2}{E_{thr}^2}-1\right)\ln\left(1-\frac{E_{thr}}{E}\right)
\right]\,-\,\frac{E_{thr}^2}{32\pi^2(pk)}\nn
\eeqn

In the latter equation, the vector $P'^\mu$ is defined as 
$P'^\mu=P^\mu-\frac{PK}{K^2}K^\mu$ with $P'^2=P^2-\frac{(PK)^2}{K^2}$, where 
$K^\mu={1\over2}(k+k')^\mu$.

Finally, the integral $I_4$ is given by
\beqn
I_4&=&\int\frac{|\vec{k}_1|^2d|\vec{k}_1|d\Omega_{k_1}}{2E_1(2\pi)^3}
\frac{(Q_1^2-Q_2^2)^2}{Q_1^2Q_2^2}\,=\,16L_\mu L_\nu J_2^{\mu\nu}\nn\\
&=&8L^2(2L^2J_{24}+2J_{25})\,=\,2t(B_{11}-B_{12})\nn\\
&=&\frac{-t}{16\pi^2}\frac{E_{thr}^2}{E^2}
\left[\ln\left(\frac{4E_{thr}^2}{m^2}\right)-3\right].
\eeqn

\end{document}